\documentclass{caosp306}

\usepackage{graphicx}

\usepackage{natbib}
\articleNo{1}
\pubyear{2019}
\volume{49}
\volnumber{1}
\firstpage{85}
\received{December 2, 2018}
\accepted{January 30, 2019}

\def\BibTeX{{\rm B\kern-.05em{\sc i\kern-.025em b}\kern-.08em
             T\kern-.1667em\lower.7ex\hbox{E}\kern-.125emX}}

\begin{document}

%
\hauthor{L.\,Zampieri {\it et al.}}

\title{(Very) Fast astronomical photometry for meter-class telescopes}


%
\author{
        L.\,Zampieri \inst{1}
      \and 
        G.\,Naletto \inst{2}
 \and 
        C.\,Barbieri \inst{1,2}
 \and 
        A.\,Burtovoi \inst{1,2}
 \and 
        M.\,Fiori \inst{2}
 \and
        A.\,Spolon \inst{2}
 \and 
        P.\,Ochner \inst{2}
 \and 
        L.\,Lessio \inst{1}
 \and 
        G.\,Umbriaco \inst{2}
 \and 
        M.\,Barbieri \inst{3}
        }

%
\institute{
           INAF-Astronomical Observatory of Padova\\
           Italy, \email{luca.zampieri@inaf.it}
         \and
           University of Padova
         \and
           University of Atacama
           }

\date{December 1, 2018}

\maketitle

\begin{abstract}
Our team at the INAF-Astronomical Observatory of Padova and the University of Padova is engaged in the design, construction and operations of instruments with very high time accuracy in the optical band for applications to High Time Resolution Astrophysics and Quantum Astronomy. Two instruments were built to perform photon counting with sub-nanosecond temporal accuracy, Aqueye+ and Iqueye. Aqueye+ is regularly mounted at the 1.8m Copernicus telescope in Asiago, while Iqueye was mounted at several 4m class telescopes around the world and is now attached through the Iqueye Fiber Interface at the 1.2m Galileo telescope in Asiago. They are used to perform coordinated high time resolution optical observations and, for the first time ever, experiments of optical intensity interferometry on a baseline of a few kilometers. I will report on recent technological developments and scientific results obtained within the framework of this project.
\keywords{Astronomical instrumentation, methods and techniques -- Instrumentation: photometers -- Techniques: interferometric -- Occultations -- pulsars: general -- pulsars: individual: PSR J0534+2200 -- X-rays: binaries -- X-rays: individual: MAXI J1820+070
}
\end{abstract}

\section{Introduction}
\label{intr}

The origin of the AQUEYE+IQUEYE project\footnote{https://web.oapd.inaf.it/zampieri/aqueye-iqueye/index.html} dates back to September 2005, when we completed an instrument design study (QuantEYE, the ESO Quantum Eye; \citealt{2005astro.ph.11027D}) within the framework of the proposal for new instrumentation for the 100 m diameter Overwhelmingly Large (OWL) telescope. The main objective of the study was to demonstrate the possibility to reach picosecond time resolution in the optical band, needed to bring Quantum Optics concepts into the astronomical domain. We had two main scientific goals in mind: measuring the entropy of light through the statistics of the photon arrival times and performing optical High Time Resolution Astrophysics with sub-ms time resolution.

Starting from the QuantEYE design, we realized two similar instruments, Aqueye+ and Iqueye. They are non-imaging photon-counting instruments dedicated to performing very fast optical photometry, with a field of view of few arcsec and the capability of time tagging the detected photons with sub-ns time accuracy. This instrumentation is extremely versatile because the photon event lists are stored in a mass memory and the data analysis is entirely done in post-processing with a selection of time bins (from nanoseconds to minutes). In addition, simultaneous observations with distant telescopes can be made, because a suitable synchronization of the signals is implemented. This fact allows us to correlate the simultaneous acquisitions from the two instruments.

The first instrument that was realized is Aqueye, the Asiago Quantum Eye, mounted on the AFOSC instrument at the Copernicus telescope in Asiago \citep{2009JMOp...56..261B}. The second step was Iqueye, the Italian Quantum Eye, designed for applications to 4 m class telescopes \citep{2009A&A...508..531N}. It was mounted on the ESO 3.5 m New Technology Telescope in La Silla, and the Telescopio Nazionale Galileo and the William Herschel Telescope in La Palma.

In 2014 Aqueye become an independent instrument, called Aqueye+ and no longer attached to AFOSC \citep{2013SPIE.8875E..0DN}. It is now remotely controlled and regularly mounted at the 1.8 m Copernicus telescope in Asiago \citep{2015SPIE.9504E..0CZ}. The acquisition electronics was moved to a dedicated room and is kept under stable temperature and humidity conditions. In 2015 Iqueye returned to Asiago for a general refurbishment and was mounted at the 1.2 m Galileo telescope to perform experiments of intensity interferometry \citep{2016SPIE.9907E..0NZ}. A direct mount at the Cassegrain focus was soon discarded because of potential mechanical problems. We opted for a soft-mount solution, installing a dedicated optical bench at the Nasmyth focus of the Galileo telescope and connecting Iqueye to it through an optical fiber. 

Since 2015 the two instruments are used to perform coordinated high time resolution optical observations and, for the first time ever, exploratory experiments of stellar intensity interferometry on a baseline of a few kilometers. Some of the results obtained within the framework of these programs are reported here.

\section{Instrumentation}
\label{aqueye+iqueye}

Both instruments, Aqueye+ and Iqueye, adopt a very convenient optical design (\citealt{2009JMOp...56..261B,2009A&A...508..531N}). The detectors are Single Photon Avalanche Photodiodes (SPADs), photon counting detectors with high quantum effciency ($\sim 50$\%) and very high time resolution (30-50 ps) but rather small effective area. To better couple the small detectors effective area to the telescope pupil and to increase the maximum sustainable count rate, Aqueye+ and Iqueye implement the same concept of splitting the telescope entrance pupil proposed for QuantEYE. The pupil is divided in four parts, each of them focused on a single detector.

The core of the instrumentation is its very accurate acquisition and timing system. The instruments time tag and store the arrival time of each detected photon with a $\simeq 100$ ps relative time accuracy and $< 500$ ps absolute time accuracy (compared to UTC; \citealt{2009A&A...508..531N}). All recorded times are stored in event lists that can be analyzed in post-processing (\citealt{2015SPIE.9504E..0CZ}). At present the most important limitation is the maximum data rate, of the order of few MHz (in the linear regime).

\begin{figure}
\centerline{\includegraphics[width=0.75\textwidth,clip=]{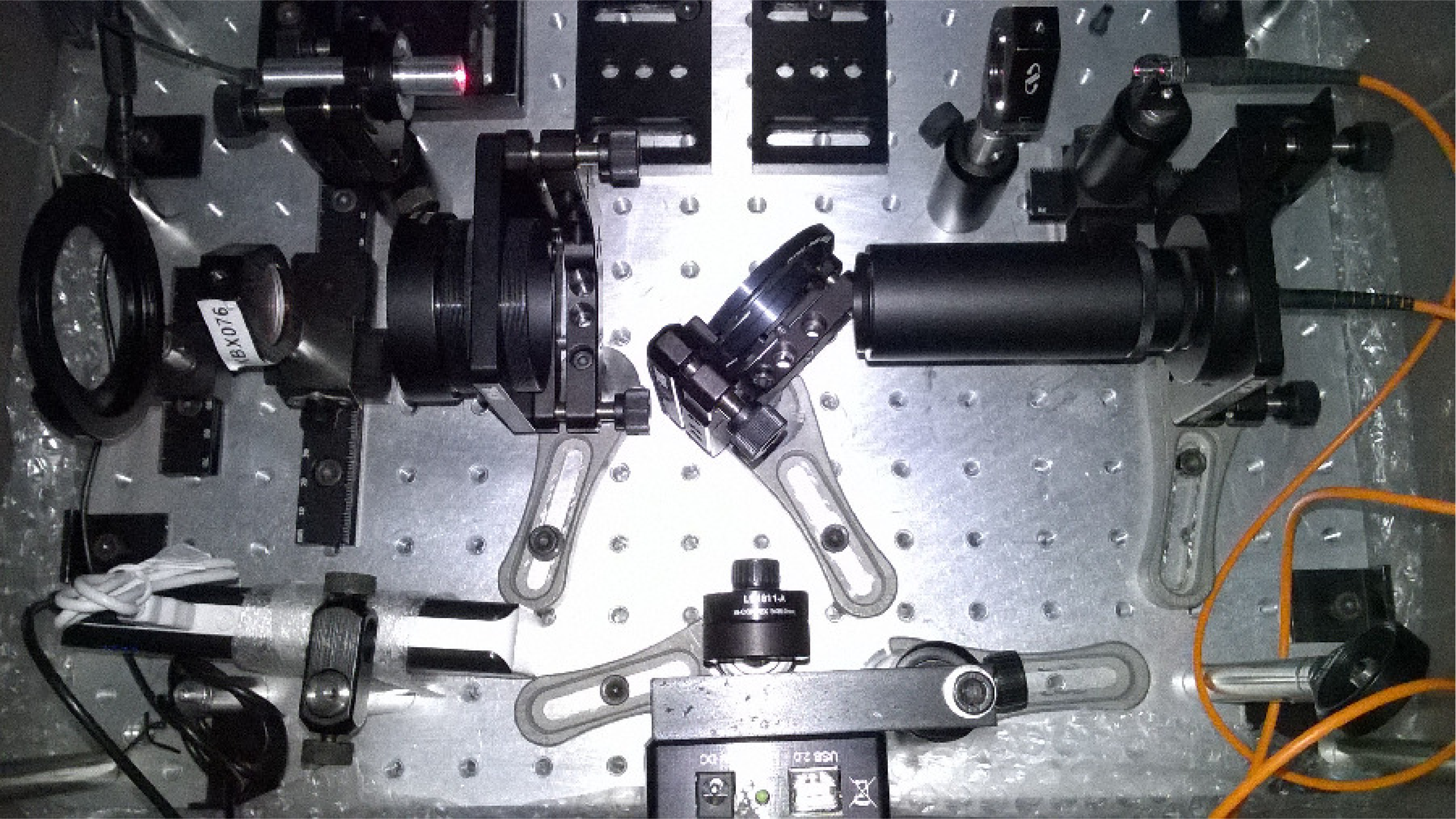}}
\centerline{\includegraphics[width=0.9\textwidth,clip=]{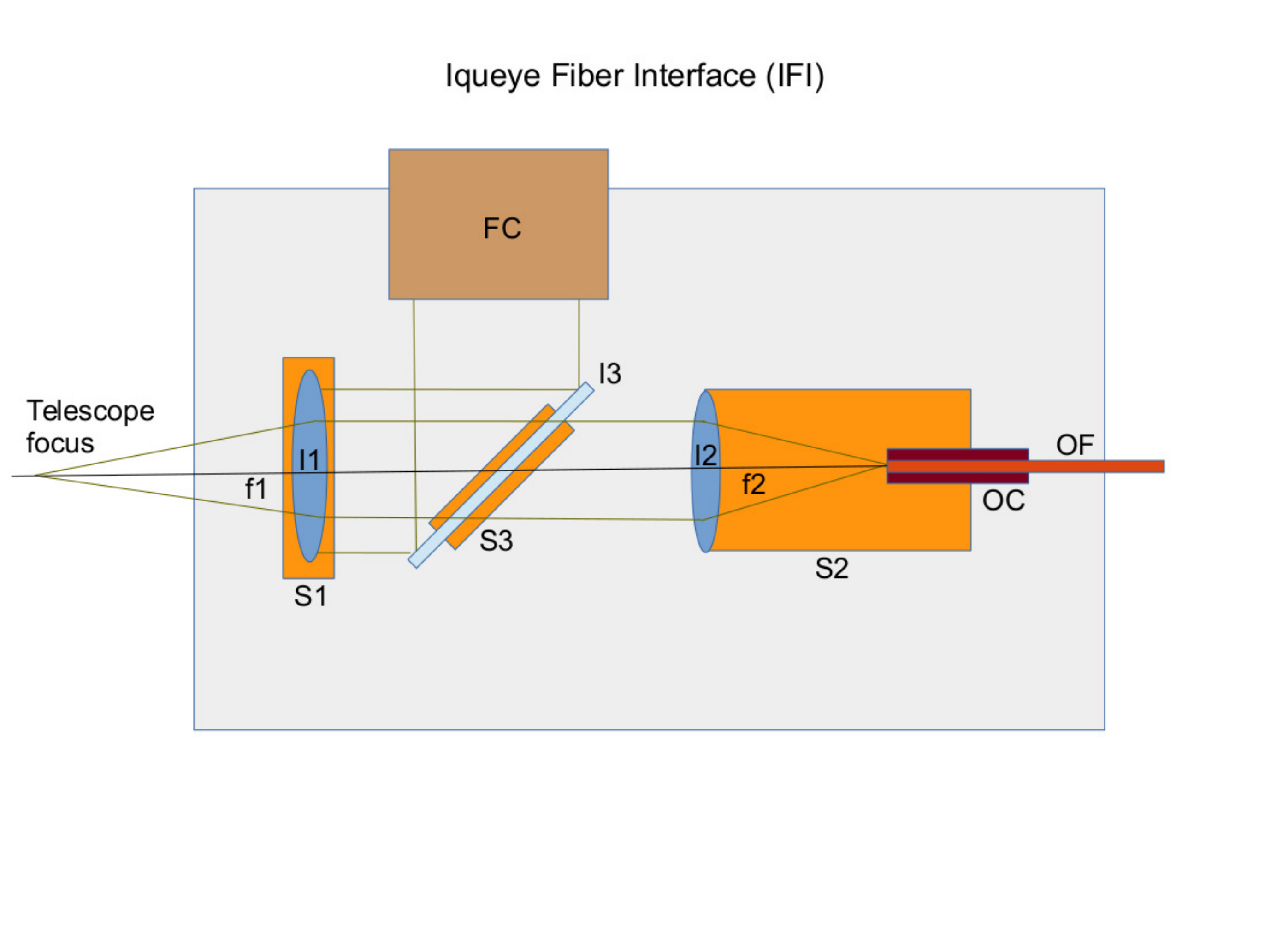}}
\caption{{\it Top}: Iqueye Fiber Interface (IFI). Opto-mechanical interface for coupling Iqueye with the 1.2 m Galileo telescope, attached at the Nasmyth focus. {\it Bottom}: Optical design and main opto-mechanical components of IFI. After the telescope focus the incoming beam is collimated through an achromatic lens doublet (I1) and then focused on the optical fiber (OF) with a second achromatic doublet (I2). A beam splitter (I3) is inserted in the collimated portion of the beam. S1, S2 and S3 are the corresponding opto-mechanical supports, and OC is the fiber connector. I3 reflects 8\% of the incoming light towards a field camera (FC), and transmits the remaining 92\% to the optical fiber. The focal lengths of the two doublets are $f_1=200$ mm and $f_2=100$ mm, leading to an overall demagnification of 1:2. The core of the (multimode) optical fiber has a diameter of 365 micrometers, that corresponds to 12.5 arcsec. The image of the field camera has a plate scale of 62.3 arcsec/mm and a field of view of 11.8 x 7.4 arcmin$^2$. Filters can be inserted in the collimated beam between I1 and I3.}
\label{fig:ifi}
\end{figure}

\section{Interfacing the instruments with optical fibers}
\label{fiberinterfaces}

In order to connect Iqueye to the 1.2 m Galileo telescope, we were forced to opt for a soft-mount solution. To avoid potential problems related to the mechanics of the telescope, we installed a dedicated optical bench at the telescope lateral focus and connected Iqueye to it by means of an optical fiber \citep{2016SPIE.9907E..0NZ}. Besides facilitating the mounting of the instrument, this solution has also the advantage of maintaining Iqueye in a separate room under controlled temperature and humidity conditions (reducing potential systematics related to varying ambient conditions). At the same time, it mitigates scheduling requirements related to the time needed to mount and dismount the instrument. 

Since 2015, the bench was modified and upgraded, becoming an indepedent instrument, the Iqueye Fiber Interface (IFI; Figure~\ref{fig:ifi}).  The alignment, focusing and overall optical efficiency of IFI (including the optical fiber) have been tested in the laboratory. Taking into account the loss on the beam splitter (that re-directs 8\% of the light to the field camera) the total transmittance efficiency is $\sim 80$\%. At present, IFI can be mounted and used independently for fiber-coupling also other instruments and/or performing imaging and photometry with the field camera.

Building on the experience gained with IFI, a parallel work for the implementation of an optical fiber interface is under way also for Aqueye+. In this case, the main purpose is to trigger a prompt use of Aqueye+ in ToO mode for targeting transient or short duration events. As for Iqueye, this solution has also the advantage of maintaining Aqueye+ in a dedicated, thermally controlled room in the telescope dome. The final solution adopted for fiber-coupling Aqueye+ is different from that of Iqueye, because the Nasmyth focus of the Copernicus telescope is not available. After considering other possibilities \citep{2016SPIE.9907E..0NZ}, eventually we opted for positioning an optical fiber directly at the telescope focal plane, exploiting the mechanical support provided by AFOSC and the telescope field camera for pointing and guiding. This activity is still ongoing.

For both instruments, a major concern is the efficiency of the optical coupling. In this respect, a crucial problem to face is to minimize the losses when injecting light into the instruments. To this end, one needs to carefully select the properties of the optical fiber and to insert additional optical elements. The light injection from the fiber into each instrument is realized by means of a dedicated opto-mechanical module (module Z), that acts as a focal multiplier. A module is placed in front of each instrument, and is properly centered and focussed. For Iqueye the magnification of this module is 2.5, while for Aqueye+ it is 2.

\section{Scientific goals, experiments and projects}
\label{science}

As mentioned above, one of the main goals of the AQUEYE+IQUEYE project is achieving tens of ps time resolution, needed to bring Quantum Optics concepts into the astronomical domain. We are pursuing it in exploratory experiments, in particular the measurement of the second order correlation of star light through a modern version of the Hanbury Brown and Twiss Intensity Interferometry experiment (\citealt{1974MNRAS.167..121H} and references therein). In 2016 we started {\it an experiment of stellar intensity interferometry on} a km baseline using the Asiago telescopes and our instrumentation \citep{2016SPIE.9907E..0NZ,2016SPIE.9980E..0GN}. 
Nowadays, there is renewed interest for this technique because of the possibility of exploiting multiple and very long (km) baselines
{\it to image and resolve stellar environments} on sub-milli-arcsec angular scales. 

The other main goal of the project is performing optical High Time Resolution Astrophysics with ms or sub-ms time resolution to study rapidly varying phenomena in different types of astrophysical sources. We are pursuing it in a number of regular observing programs tailored to:
\begin{itemize}
\item Optical pulsars (in synergy with facilities operating at other wavelengths), to study with unprecedented accuracy their timing behaviour and pulse shape
\item Timing of optical transients and counterparts of binary systems with compact objects, to search for rapid optical variability
\item Lunar occultations (in synergy with other telescopes in Europe), to measure stellar radii/asymmetries with milli-arcsec angular resolution and determine the binary nature and separation of tight binary systems
\item Transits/occultations of Trans-Neptunian Objects, to determine their orbit and physical properties
\end{itemize}

In the following we will report on recent results obtained from the first two programs.

\begin{figure}
\centerline{\includegraphics[width=0.75\textwidth,clip=]{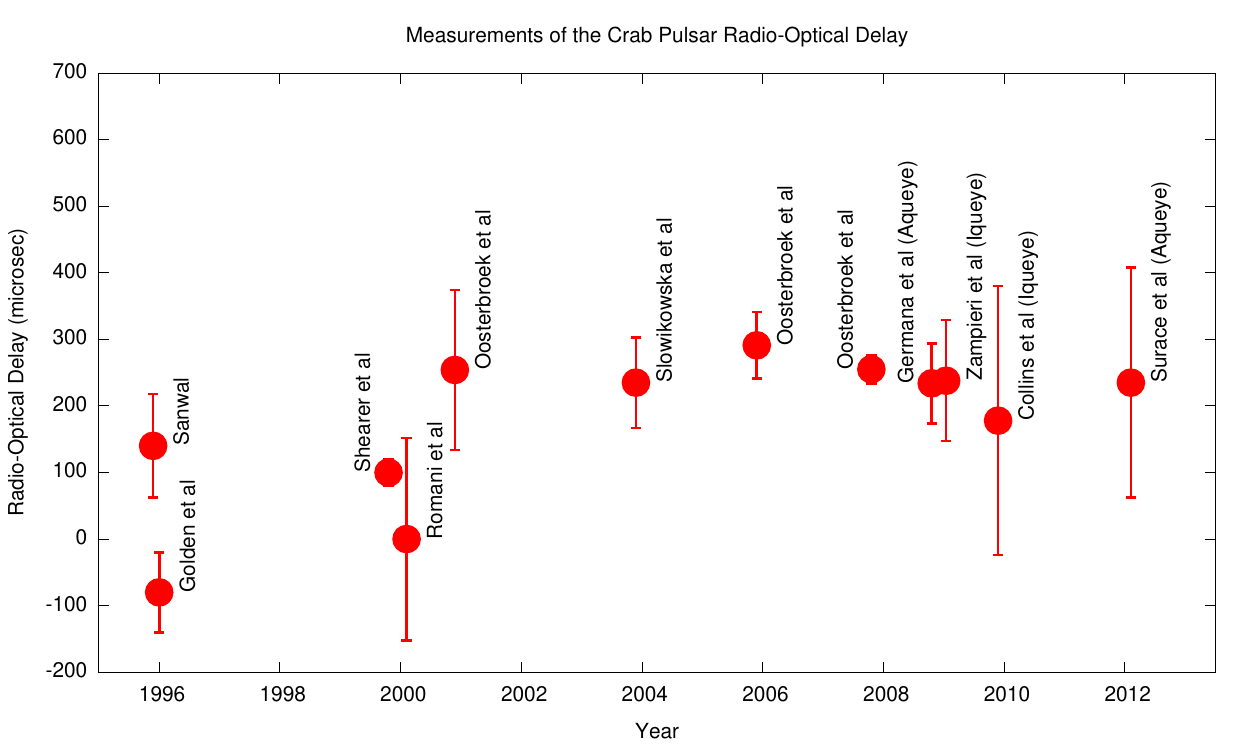}}
\caption{Delay between the time of arrival of the main pulse of the Crab pulsar in the radio band and that in the optical band. Different points refer to measurements taken by different Authors (as indicated) with different instruments since 1996. Results before year 2000 may be affected by significant systematic errors. We are regularly monitoring this delay since 2008.}
\label{fig:radoptdelay}
\end{figure}

\section{Scientific results}
\label{results}

\subsection{Very fast photometry (timing) of optical pulsars}
\label{pulsars}

Within this program we perform a regular monitoring of the Crab pulsar with both Aqueye+@Copernicus (e.g. \citealt{2012A&A...548A..47G}) and IFI+Iqueye@Galileo. At present, this is the only regular optical monitoring program of the Crab pulsar and represents the optical analogous of the Jodrell Bank radio monitoring program\footnote{http://www.jb.man.ac.uk/~pulsar/crab.html}. Observations of the Crab pulsar were taken also with Iqueye@NTT in two runs (Jan, Dec) in 2009 \citep{2014MNRAS.439.2813Z}, when very fast photometry (timing) of a number of optical pulsars including PSR B0540-69 \citep{2015Sci...350..801F} and the Vela pulsar \citep{2019MNRAS.482..175S}, was carried out.

One of the main goals of this program is to study the evolution of the delay between the time of arrival of the main pulse in the radio band and that in the optical band. It is quite well established that, since at least the beginning of 2000, the pulse in the optical leads that in the radio by 150--250 $\mu$s (e.g. \citealt{2008A&A...488..271O} and references therein). Our most recent measurements, taken since 2008, confirm these findings (Figure~\ref{fig:radoptdelay}; see \citealt{2012A&A...548A..47G,2014MNRAS.439.2813Z}). The optical and radio beams are probably misaligned (1.5--3 deg) because at the position where electrons emit optical photons the magnetic field has a slightly different orientation. Regular monitoring observations are carried out to search for a possible evolution of this delay, that may be induced by a significant reconfiguration of the magnetic field of the pulsar or by a change in the geometry of the emission region.

Besides a delay, the optical timing solution calculated on a timescale of a few days shows a significant drift from the radio ephemeris. This difference may be related to the optical measurements tracking the intrinsic daily/weekly pulsar noise \citep{2016A&A...587A..99C} and/or to dispersion measure variations not completely removed from radio data (while optical data do not depend on them). In this respect, optical data provide a robust and independent confirmation of the radio timing solution and a means to study the ``typical'' Crab pulsar frequency noise on a daily timescale. Another reason to monitor the timing history of the Crab pulsar in the optical band is to study its short term evolution after the occurrence of glitches and/or flares in the Crab nebula (e.g. \citealt{2016A&A...587A..99C}).

\subsection{Very fast photometric monitoring of optical transients}
\label{transients}

MAXI 1820+070 is a bright and uncatalogued X-ray transient source discovered on Mar 11, 2018 by \cite{2018ATel11399....1K}, later identified with the optical transient ASASSN-18ey. The source was proposed to be a candidate black hole X-ray binary by \cite{2018ATel11418....1B}. It shows pronounced and fast variability/flaring activity in both the X-ray and optical bands \citep{2018ApJ...867L...9T}.

After the discovery of the source, in April 2018 we promptly started a fast-photometry monitoring campaign with IFI+Iqueye and Aqueye+ \citep{2018ATel11723....1Z,2018ATel11936....1Z,2018ATel11824....1F}. This effort was carried out in synergy with the SMARTNet multiwavelength network\footnote{https://www.isdc.unige.ch/SMARTNet/}. A joint photometric and spectroscopic campaign was simultaneously undertaken with the Asiago and ANS Collaboration telescopes \citep{2018ATel12157....1M,2018ATel11899....1M}.
We found that, compared to a nearby reference star, in Apr and Jun 2018 the power spectrum of MAXI J1820+070 shows significant red noise and quasi periodic oscillations (QPOs).

\begin{figure}
\centerline{\includegraphics[width=0.75\textwidth,clip=]{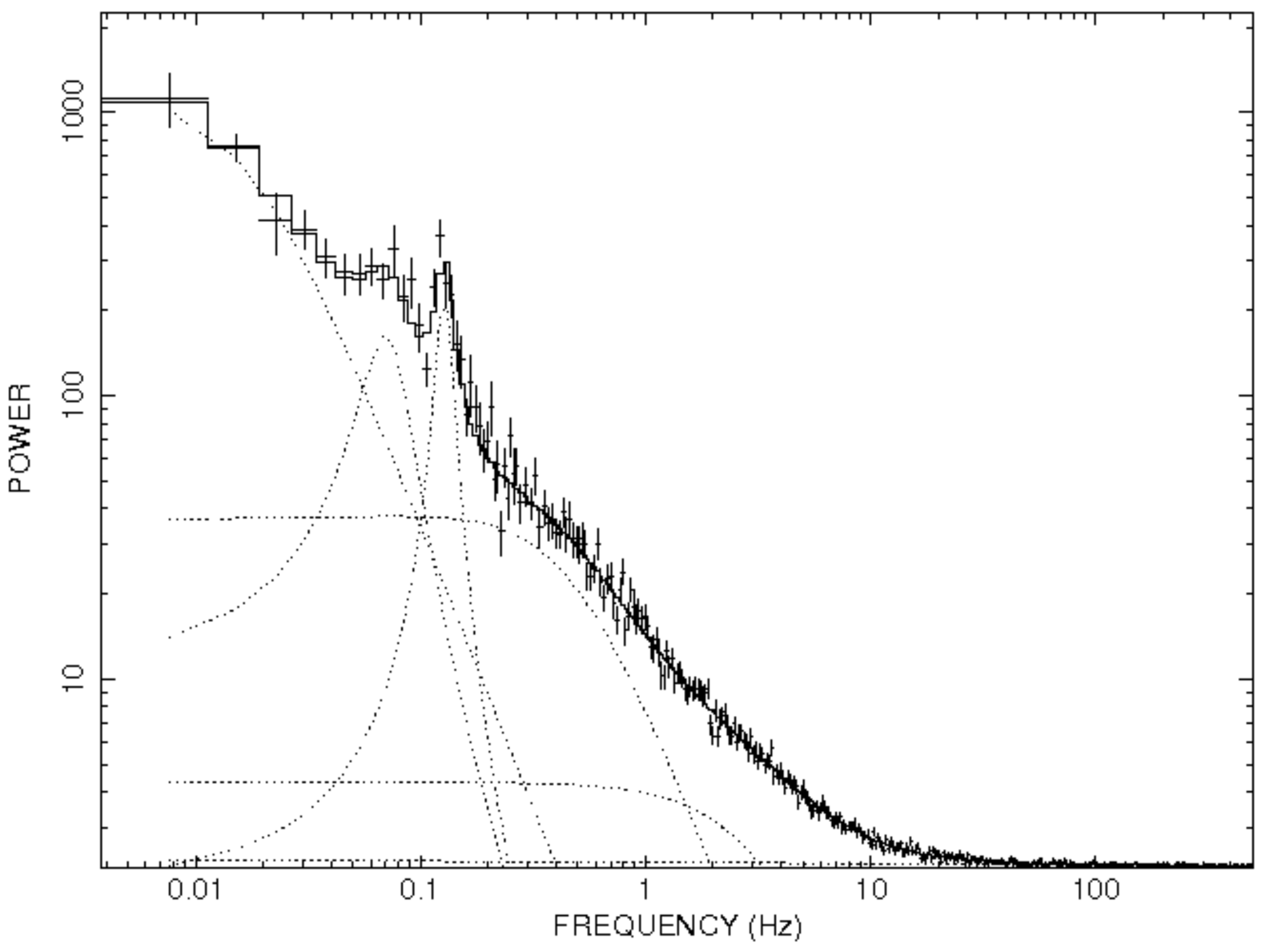}}
\caption{Power density spectrum (Lehay normalized) of MAXI 1820+070 taken with IFI+Iqueye on Apr 18-19, 2018 \citep{2018ATel11723....1Z}. Four observations for a total duration of 3600s were considered. Power spectra were computed from the non-background subtracted light curves with a time bin of 1 ms and were then averaged over intervals of 130s duration.}
\label{fig:maxipds}
\end{figure}

The power spectrum of 3600s of data taken on Apr 18-19, 2018 is shown in Figure~\ref{fig:maxipds}. We detect a significant quasi-periodic oscillation (QPO) on the top of three broad-band noise components (in part induced by the sky background; \citealt{2018ATel11723....1Z}). The QPO has a centroid frequency of 128$\pm$2 mHz, a full-width-half-maximum of 24$\pm$5 mHz, and a fractional root-mean-square (rms) variability of 3.1$\pm$0.3\%. Another QPO-like feature at lower frequency and with lower significance is present in the power spectrum with centroid frequency 71$\pm$4 mHz, full-width-half-maximum 36$\pm$16 mHz, and fractional rms variability 3.2$\pm$0.6\%. The properties (frequency and width) of the 128 mHz optical QPO are also consistent with those of the QPO at $\sim$0.12 Hz detected in the power spectrum of quasi-simultaneous Swift observations of MAXI J1820+070 taken between Apr 16 and 20, 2018. Furthermore, the 128 mHz QPO has a width consistent with that of the optical QPO reported in \cite{2018ATel11591....1Y}. The frequency difference between the two measurements, that are approximately 3.7 days apart, is 29 mHz. The fractional increment of the centroid frequency is consistent with that calculated for a similar QPO detected in the X-rays with {\it NuSTAR} and reported in \cite{2018ATel11578....1B}.

On June 9-10, 2018, we detected other two QPO-like features \citep{2018ATel11824....1F}. The power spectra of 4500s of data, calculated with a time bin of 1 ms and then averaged over intervals of 65s duration, show two broad peaks on the top of a broad-band noise component. The higher frequency QPO-like feature has a centroid frequency of 268$\pm$12 mHz, a full-width-half-maximum of 150$\pm$39 mHz, and a fractional rms variability of 1.9$\pm$0.2\%. The lower frequency QPO is less significant and has a centroid frequency of 151$\pm$6 mHz, a full-width-half-maximum of 33$\pm$16 mHz, and a fractional rms variability of 1.2$\pm$0.3\%. We obtained acceptable fits of the power spectrum also with two harmonically-related QPOs, that have width and significance similar to those reported above. Acceptable fits are obtained for 1:2, 2:3, or 3:5 centroid frequency ratios. A fit with harmomically-related QPOs performed on the April IFI+Iqueye observations gives similar results for the same harmonic ratios (1:2, 2:3 or 3:5).

The origin of the low frequency X-ray QPOs detected in black hole X-ray binaries has been matter of intense debate. They are thought to originate from the Lense–Thirring (LT) precession of the hot inner part of the accretion disc, tilted from the equatorial plane of the central rotating black hole \citep{2009MNRAS.397L.101I}. The recent detection of QPOS in the optical and infrared band, in addition to the X-ray band, provides precious additional information to understand this phenomenon. Optical and infrared QPOs may be produced directly from the inner disc through synchrotron emission, modulated at the LT precession frequency \citep{2015MNRAS.454.2855V}. Or, the inner disc undergoes LT precession and illuminates varying outer regions of the accretion disc, generating a reprocessed optical QPO \citep{2015MNRAS.448..939V}. If the optical/infrared emission comes from a jet launched from the accretion disc, jet precession/modulation may be driven by LT precession of the disc \citep{2018MNRAS.480.2054M}.

\section{Conclusions}
\label{conclusions}

Aqueye+ and IFI+Iqueye are two very fast photon counters operating in Asiago.
They are used for performing very fast photometry of astrophysical sources and modern implementation experiments of the Hanbury Brown and Twiss Intensity Interferometer. Iqueye is fiber-fed through an independent instrument (IFI).

As part of our very-fast photometry observing programs, we are regularly monitoring the Crab pulsar radio-optical delay since 2008.
No significant variation of the delay has occurred, indicating that the relative position and geometry of the radio and optical emission regions have not changed.
We carried out also a monitoring campaign of the X-ray transient MAXI J1820+070,
detecting a number of QPOs in the power spectrum that will provide precious additional information on this phenomenon.
We plan to continue the presently active experiments and observing programs, and to make Aqueye+@Copernicus fiber-fed and permanently available in ToO mode.
We may in principle export this technique for coupling instruments at other meter-class facilities.

Very fast optical multicolor photometry allows a variety of prime interest scientific questions to be addressed in the domain of High Time Resolution Astrophysics even with meter-class telescopes. Coordinated multi-wavelength campaigns most often benefit from the support of small telescopes world-wide to guarantee adequate temporal coverage. At the same time, forefront instrumentation allows for unique observations to be made. Highly experimental programs, technological development activities, and training of young astronomers can be run almost exclusively on such facilities.

\acknowledgements
We are grateful to all the staff of the Asiago observing station (Venerio Chiomento, Giancarlo Farisato, Aldo Frigo, Giorgio Martorana, Luciano Traverso, Robertino Bau', Giovanni Costa, Alessandro Siviero) for their work and support. We also thank Marco Fiaschi, Tommaso Occhipinti, Enrico Verroi and Paolo Zoccarato for their external support to the project. This wok is based on observations collected at the Copernicus telescope (Asiago, Italy) of the INAF-Osservatorio Astronomico di Padova and at the Galileo telescope (Asiago, Italy) of the University of Padova. This project is partly supported by the University of Padova under the Quantum Future Strategic Project, by the Italian Ministry of University MIUR through the programme PRIN 2006, by the Project of Excellence 2006 Fondazione CARIPARO, by INAF-Astronomical Observatory of Padova under the grant ``Osservazioni con strumentazione astronomica ad elevata risoluzione temporale e modellizzazione di emissione ottica variabile'', and by Fondazione Banca Popolare di Marostica-Volksbank.

\bibliography{proceedings_aqueye+iqueye_project_slovakia_2018_proc_v4_arxiv}
\end{document}